# Plasmonic Parametric Resonance

Alessandro Salandrino[1,2*]

[1]*University of Kansas, Department of Electrical Engineering and Computer Science*
*1520 West 15th Street, Lawrence, KS, USA 66045*
[2]*University of Kansas, ITTC*
*2335 Irving Hill Road, Lawrence, KS, USA 66045*

[*]*e-mail address: a.salandrino@ku.edu*

We introduce the concept of Plasmonic Parametric Resonance (PPR) as a novel way to amplify high angular momentum plasmonic modes of nanoparticles by means of a simple uniform optical pump. In analogy with parametric resonance in dynamical systems, PPR originates from the temporal modulation of one of the parameters governing the evolution of the state of the system. As opposed to conventional localized surface plasmon resonances, we show that in principle any plasmonic mode of arbitrarily high order is accessible by PPR with a spatially uniform optical pump. Moreover, in contradistinction with other mechanisms of plasmonic amplification, the coherent nature of PPR lends itself to a more straightforward experimental detection approach. The threshold conditions for PPR are analytically derived. Schemes of experimental realization and detection are also discussed.

Localized Surface Plasmons (LSP) are non-propagating coherent oscillations of free-carriers confined in plasmonic particles [1]. These modes can be externally excited by photonic or electronic scattering [2], leading to strongly localized electric fields in proximity of the particle's surfaces. An enhanced optical response is obtained when LSPs are resonantly excited by an incident field at the characteristic frequency of the dipolar eigenmode. In addition to their dipolar response plasmonic particles, in general, support an infinite discrete set of plasmonic resonances [3, 4] associated with high electromagnetic angular momentum states. In the simple case of a plasmonic nanosphere with frequency dispersive permittivity $\varepsilon_1(\omega)$ embedded in a background medium with constant permittivity $\varepsilon_2$, for a resonance of order $n \geq 1$ there are $2n+1$ degenerate angular-momentum states with complex frequency $\omega_n$ satisfying the condition $\varepsilon_1(\omega_n) = -(1+n)\varepsilon_2/n$. For $n \gg 1$ the eigenmodes tend to occur for $\varepsilon_1(\omega_{n \gg 1}) \sim -\varepsilon_2$. The increased modal density for $\varepsilon_1 \sim -\varepsilon_2$ is not exclusive to spherical particles, but is rather a general feature of all plasmonic structures [5]. Accessing such spectrally dense set of tightly bound resonant modes would lead to enhanced nonlinear light-matter interactions at the nanoscale, with applications ranging from sensing [6, 7] and Raman spectroscopy [8, 9] to near-field nonlinear optics [10, 11], imaging [12] and nano-manipulation [13, 14], as well as for the realization of optical metamaterials [15].

The efficiency with which such resonances can be excited by an external incident field depends upon the spatial and spectral overlap between the excitation field and the specific plasmonic mode, or in other words on how closely the conditions of energy and angular momentum conservation are met. For deeply subwavelength plasmonic particles only the lowest order mode of electric dipolar nature is efficiently coupled to, and excited by, radiation states. The higher order eigenmodes tend to be subradiant, and by reciprocity they are nearly decoupled from free-space propagating fields. Therefore, exciting such higher order modes requires either sophisticated near-field scattering techniques [16], or the use of active media to promote surface plasmon amplification by stimulated emission of radiation (SPASER) [17]. Just as challenging is the optical detection of such modes, due to their nearly non-radiating nature.

The aforementioned limitations of LSP are largely dependent upon the optical excitation mechanism. To overcome these limitations we present an alternative approach to deliver energy to the free-carriers of a plasmonic nanoparticle by introducing a different form of LSP resonance: the plasmonic parametric resonance (PPR). PPR originates from the temporal modulation of one of the parameters governing the evolution of the state of the system. In this Letter we outline the theory of PPR in comparison with conventional LSP resonance. In particular, we show that in principle any plasmonic mode is accessible by PPR using a *pump field that is spatially uniform*, thereby overcoming the difficulty of matching the spatial profile of high-angular momentum modes. An example of a possible experimental realization is discussed and analyzed both theoretically and numerically, showing that PPR can be achieved and observed in a realistic system.

For the purpose of illustration of PPR we consider a system which is amenable to a closed-form solution: a subwavelength plasmonic sphere in a homogeneous dielectric background medium. More complex configurations would display qualitatively similar phenomenology. We consider a sphere of radius $R$ and relative permittivity $\varepsilon_1$ (medium 1), embedded in a uniform dielectric medium $\varepsilon_2$ (medium 2). The radius $R$ is assumed to be much smaller than the free-space wavelength associated with any of the plasmonic eigenmodes of interest, such that a quasi-static approach is applicable for determining the spatial distribution of the electromagnetic field. The dispersion of $\varepsilon_2$ is neglected. Medium 1 is assumed to follow a Drude-like frequency-domain dispersion $\varepsilon_1(\omega) = \varepsilon_\infty - \omega_{pl}^2/(\omega^2 + i\omega\gamma)$, with plasma frequency $\omega_{pl}$, collision frequency $\gamma$, and a non-dispersive term accounting for high-frequency spectral features, $\varepsilon_\infty$. The dispersive term in the $\varepsilon_1(\omega)$ expression is associated with the equation of motion for the free-carrier polarization density $\mathbf{P}_1(\mathbf{r},t)$ within medium 1:

$$\partial_t^2 \mathbf{P}_1 + \gamma_{tot} \partial_t \mathbf{P}_1 = \omega_{pl}^2 \varepsilon_0 \mathbf{E}_1 \qquad (1)$$

In equation (1) the damping rate $\gamma_{tot}$ is corrected to include radiation effects in addition to the collision frequency $\gamma$, i.e. $\gamma_{tot} = \gamma + \gamma_{rad}$. $\mathbf{E}_1$ is the electric field in the region occupied by medium 1.

In the quasi-static limit the potential of an electric multipole eigenmode of order $(n,m)$ can be expressed in terms of even and odd spherical harmonics $Y_{n,m}^{(e/o)}(\phi,\theta)$. Under these assumptions the polarization density profile $\mathbf{P}_1(\mathbf{r},t)$ can be expressed as superposition of spherical harmonics with time-varying coefficients $P_{n,m}^{(e/o)}(t)$:

$$\mathbf{P}_1(\mathbf{r},t) = \sum_{n=0}^{\infty}\sum_{m=0}^{n}\sum_{j=e,o} P_{n,m}^{(j)}(t) \nabla\left[\frac{r^n}{R^{n-1}} Y_{n,m}^{(j)}(\theta,\phi)\right] \qquad (2)$$

Exploiting the orthogonality of spherical harmonics, the application of the electromagnetic boundary conditions at $r = R$ yields separate equations of motion for the polarization density amplitudes $P_{n,m}^{(e/o)}(t)$ of each angular momentum state:

$$\frac{d^2 P_{n,m}^{(e/o)}(t)}{dt^2} + \gamma_{n,m}\frac{dP_{n,m}^{(e/o)}(t)}{dt} + \Omega_n^2 P_{n,m}^{(e/o)}(t) = 0 \qquad (3)$$

where $\Omega_n^2 = \{\omega_{pl}^2 n/[\varepsilon_2 + n(\varepsilon_2 + \varepsilon_\infty)]\}$ is the modal frequency of the plasmonic eigenmode of order $(m,n)$ in the absence of damping, and $\gamma_{n,m} = \gamma + \gamma_{n,m}^{rad}$ is the damping rate including material absorption effects $\gamma$ and modal radiation effects $\gamma_{n,m}^{rad}$. For $n \gg 1$ the radiation damping becomes negligible and $\gamma_{n\gg 1,m} \to \gamma$.

The energy in the system at any point in time is partitioned and exchanged between potential energy $U_{n,m}(t)$ and kinetic energy $K_{n,m}(t)$ which can be simply expressed in terms of the initial phase-space coordinates $\left[P_{n,m}(0), \dot{P}_{n,m}(0) = (\partial_t P_n)_{t=0}\right]$. Of particular interest are the following two complementary initial conditions $\dot{P}_{n,m}(0) = 0$ or $P_{n,m}(0) = 0$:

$$\begin{cases} \dot{P}_{n,m}(0) = 0 \\ U_{n,m}(0) = \dfrac{nR^3 \Omega_n^2}{4\varepsilon_0 \omega_{pl}^2}\left[P_{n,m}(0)\right]^2 \\ K_{n,m}(0) = 0 \end{cases} \qquad (4)$$

$$\begin{cases} P_{n,m}(0) = 0 \\ K_{n,m}(0) = \dfrac{nR^3}{4\varepsilon_0 \omega_{pl}^2}\left[\dot{P}_{n,m}(0)\right]^2 \\ U_{n,m}(0) = 0 \end{cases} \qquad (5)$$

In the situation described by equation (4) the total energy is potential energy and it depends explicitly on the modal frequency $\Omega_n$, which can be expressed in terms of the background permittivity $\varepsilon_2$. In particular the potential energy in expression (4) coincides with the energy of surface polarization charge of the $(n,m)$ component of the polarization density (2), sitting in the corresponding eigenmode electric potential $V_{n,m}^{(e/o)}(\mathbf{r},t)$:

$$V_{n,m}^{(e/o)}(r=R,\theta,\phi,t) = \frac{P_{n,m}^{(e/o)}(t)R}{\varepsilon_0 \varepsilon_2 + n\varepsilon_0(\varepsilon_2 + \varepsilon_\infty)} Y_{n,m}^{(e/o)}(\theta,\phi) \qquad (6)$$

As evident from equation (6) a decrease (increase) in the background permittivity $\varepsilon_2$ would lead to an increase (decrease) in the potential energy in the system. Assuming for the moment that $\varepsilon_2$ could be instantaneously reduced to $\varepsilon_2 - d\varepsilon_2$ (more realistic modulation conditions will be considered later), the characteristic modal frequency would change from $\Omega_n$ to $\Omega_n + d\Omega_n$. Corresponding to such increase in the modal frequency, the energy of the system increases by an amount:

$$dU_{n,m} \sim 2\frac{U_{n,m}}{\Omega_n}d\Omega_n = \left(1 + \frac{1}{n}\right)\frac{\Omega_n^2 U_{n,m}}{\omega_{pl}^2}d\varepsilon_2 \qquad (7)$$

Equation (7) indicates that the energy increment produced by the parametric modulation is proportional to the potential energy $U_{n,m}$ stored in the plasmonic mode. In the situation described in equation (5) on the other hand the system's energy is purely kinetic, and it does not depend explicitly on the modal frequency, $\Omega_n$ (or on $\varepsilon_2$). Under such conditions, an instantaneous modification of the system parameters having the sole effect of modifying the characteristic modal frequency would not affect the system's energy. From these observations, a periodic modulation scheme of the system's parameters can be identified which efficiently delivers energy to the plasmonic mode, without resorting to direct application of external electric fields on the charge carriers.

In what follows we analyze one full modulation period and determine the parametric resonance threshold for the plasmonic eigenmode of order $(n,m)$, corresponding to the regime of parametric regeneration. Let us consider a system that starts at time $t = 0$ with total energy, $W_{n,m}(0)$, in the form of potential energy of a plasmonic eigenmode of order $(n,m)$. This initial state is identified by the phase-space coordinates:

$$\begin{cases} P_{n,m}(0) = \left[ \dfrac{4\varepsilon_0 \omega_p^2}{n R^3 \Omega_n^2} W_{n,m}(0) \right]^{1/2} \\ \dot{P}_{n,m}(0) = 0 \end{cases}$$

After a time $T_1 = \left[ \pi - \tan^{-1}(2\omega_n / \gamma) \right] / \omega_n$ with $\omega_n^2 = \Omega_n^2 - \gamma^2 / 4$ the freely oscillating system evolves to a state of pure kinetic energy $K_{n,m}(T_1) = W_{n,m}(0) e^{-\gamma T_1}$ with phase space coordinates $P_{n,m}(T_1) = 0$ and $\dot{P}_{n,m}(T_1) = -P_{n,m}(0) \Omega_n e^{-\gamma T_1 / 2}$. If at this point the system's parameters are instantaneously modulated so as to modify the eigenfrequency from $\Omega_n$ to $\Theta_n = \Omega_n - d\Omega_n$, by virtue of equation (7), no energy is delivered to or taken from the system as the potential energy $U_{n,m}$ is zero at the moment. From this point on the plasmonic eigenmode oscillates at the modified eigenfrequency $\Theta_n$. The system reverts to a state of pure potential energy after a time $T_2 = (1/\theta_n) \tan^{-1}(2\theta_n / \gamma)$ with $\theta_n^2 = \Theta_n^2 - \gamma^2 / 4$, so that at time $t = T_1 + T_2 = T$ the energy of the plasmonic mode is again purely potential, with $U_{n,m}(T) = W_{n,m}(0) e^{-\gamma T}$. At time $T^+ = T + dt$ (for $dt \to 0$), if the system's parameters are instantaneously modulated to restore the eigenfrequency to the initial value $\Omega_n$, the energy of the plasmonic mode increases to a value of:

$$W_{n,m}(T^+) = W_{n,m}(0)(\Omega_n^2 / \Theta_n^2) e^{-\gamma T} \qquad (8)$$

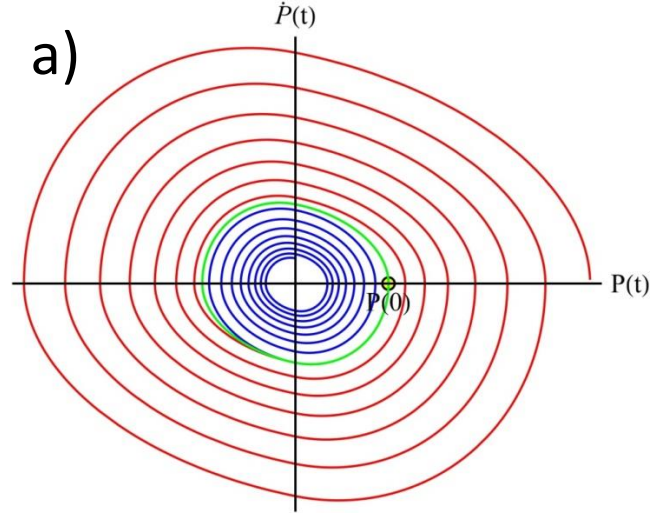

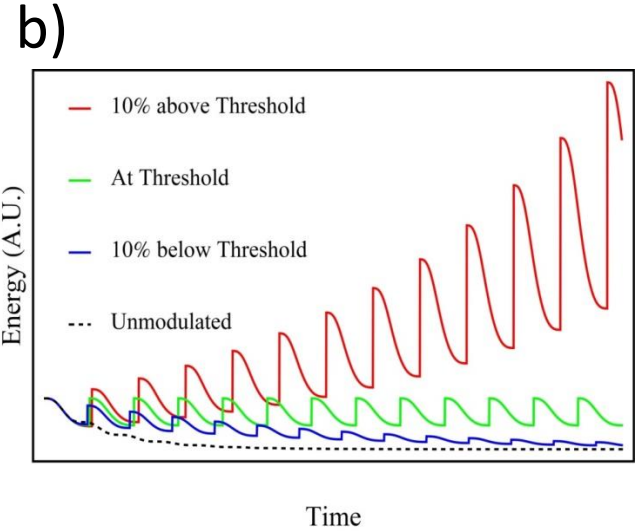

*Figure 1. a) Phase-space trajectories in terms of polarization amplitude and polarization current and b) temporal evolution of the energy content of a plasmonic mode under different parametric modulation conditions. When the PPR threshold represented by the green curve is exceeded the plasmonic mode experiences a net gain resulting in increasing polarization oscillations.*

By equating $W_{n,m}(T^+) = W_{n,m}(0)$, equation (8) yields the threshold condition for parametric resonance. In the small loss $\gamma \ll \Omega_n$ case the threshold value of the modulation depth $d\Omega_n$ can be expressed in terms of the quality factor $Q_n$ of the plasmonic mode:

$$\frac{d\Omega_n}{\Omega_n} = \frac{\pi}{2} \frac{\gamma}{\Omega_n} = \frac{\pi}{2Q_n} \qquad (9)$$

The phase-space trajectory of the state of the parametrically driven plasmonic system in terms of polarization amplitude $P_n$ and current $\dot{P}_n$ for plasmonic sphere is illustrated in Figure 1a for various modulation conditions. At the parametric regeneration threshold the phase-space trajectory is the closed orbit (i.e. the "separatrix") shown in green in Figure 1a, separating the inward and outward spiraling trajectories that occur below and above threshold respectively. The total energy of the plasmonic mode corresponding to the phase-space trajectories of Figure 1a is shown in Figure 1b. The discontinuous sections correspond to the parametric energy transfer described by equation (7) occurring when the polarization density amplitude $P_{n,m}$ is at a maximum and the polarization current density is zero. Consistent with equation (7) each parametric energy transfer is proportional to the potential energy of the polarization charges, leading to an increase in the mode energy with an exponential envelope.

The step-wise parametric modulation discussed so far is an idealization devised to determine a lower-bound for the plasmonic parametric regeneration threshold. In order to give a realistic description of an actual physical system one must consider the specific temporal profile of the modulated modal frequency $\Omega_n(t)$, which in turn depends on the physical mechanism exploited to alter the background permittivity $\varepsilon_2$. For plasmonic resonances at infrared or optical frequency, realistically, only optical nonlinearities of electronic origin would be fast enough to enable efficient PPR. As an example we analyze the practically relevant example of the harmonic modulation of the permittivity $\varepsilon_2$ mediated by its second order nonlinear optical susceptibility $\chi^{(2)}$ in the presence of a spatially uniform pump field. The inherently anisotropic character of second order nonlinear interactions [18] must be taken into account, as it introduces specific selection rules on the modes that can interact with a given pump field.

In the following we consider a plasmonic sphere (medium 1) immersed in a medium 2 belonging to the m point symmetry group [19], with a dominant second order susceptibility term $\chi_{zzz}$. Such model well describes, among others, the characteristics of 2-methyl-4-nitroanline (MNA) [20], an organic material displaying an extremely large second order susceptibility. A uniform linearly polarized pump field of the form $\mathbf{E}_P(t) = \hat{\mathbf{z}} E_P(t)$ is incident on the sphere. The interaction of the pump field with nonlinear susceptibility $\chi_{zzz}$ of medium 2 produces a nonlinear polarization density $\mathbf{P}_2^{NL} = \hat{\mathbf{z}} \varepsilon_0 \chi_{zzz} (\mathbf{E}_2 \cdot \hat{\mathbf{z}})(\mathbf{E}_2 \cdot \hat{\mathbf{z}})$. Due to the continuity of the radial component of the electric displacement, the dynamics of the polarization density in medium 1 are modified by the interaction with the radial component $\left(\mathbf{P}_2^{NL} \cdot \hat{\mathbf{r}}\right)_{r=R}$ of the nonlinear polarization in medium 2, which can be expressed as a superposition of spherical harmonics, with amplitude coefficients $S_{n,m}^{(e/o)}(t)$:

$$\left(\mathbf{P}_2^{(NL)} \cdot \hat{\mathbf{r}}\right)\bigg|_R = \sum_{n=0}^{\infty} \sum_{m=0}^{n} \sum_{j=e,o} S_{n,m}^{(j)}(t) Y_{n,m}^{(j)}(\theta,\phi) \quad (10)$$

The resulting equations of motion for the $n,m$ component of the polarization density amplitude within the sphere are given by:

$$\frac{d^2 P_{n,m}^{(e/o)}(t)}{dt^2} + \gamma \frac{dP_{n,m}^{(e/o)}(t)}{dt} + \Omega_n^2 P_{n,m}^{(e/o)}(t) = \Omega_n^2 \frac{S_{n,m}^{(e/o)}(t)}{n} \quad (11)$$

The amplitude coefficients $S_{n,m}^{(e/o)}(t)$ represent the spherical harmonic component $n,m$ of the nonlinear polarization density, originating from the mixing of various pairs of angular momentum states of the field in region 2, with selection rules dictated by the specific symmetry class of the nonlinear susceptibility under consideration. Assuming that the pump field $\mathbf{E}_P$ is much stronger than any of the plasmonic modes fields oscillating in the system – a condition valid at the onset of PPR – we shall consider only the three-wave mixing processes involving the pump. Saturation effects will occur when the electric field of the PPR mode becomes comparable in magnitude to the pump field. For the case at hand, in which the dominant component of the nonlinear susceptibility is the $\chi_{zzz}$ term, the radial component of the nonlinear polarization couples different angular momentum states, so that a number of three-wave mixing processes are allowed. In particular for a uniform $\hat{\mathbf{z}}$-polarized pump field, corresponding to a multipole of order $n_{pump} = 1$, $m_{pump} = 0$, it is tedious but straightforward to show that the interaction with a mode of order $n,m$ leads to nonlinear polarization components of order $n,m$ and $n \pm 2, m$. Considering only the angular-momentum-matched process (which is the dominant one), the leading term of the nonlinear polarization component of order $n,m$ can be expressed as:

$$\begin{cases} S_{n,m}^{(e/o)}(t) = -\left(\Omega_n / \omega_{pl}\right)^2 n u_{n,m} \chi_{zzz} E_P(t) P_{n,m}^{(e/o)}(t) \\ u_{n,m} = \dfrac{2(n+2)[n(n+1) - 3m^2]}{n[4n(n+1) - 3]} \end{cases} \quad (12)$$

The expression (12) allows for recasting equation (11) in the following more physically transparent form, in which the external permittivity modulation is represented as a parametric shift of the resonant frequency of the polarization density amplitude:

$$\begin{cases} \dfrac{d^2 P_{n,m}^{(e/o)}(t)}{dt^2} + \gamma \dfrac{dP_{n,m}^{(e/o)}(t)}{dt} + \left[\Omega_n^2 + 2\Omega_n d\Omega_{n,m}(t)\right] P_{n,m}^{(e/o)}(t) = 0 \\ d\Omega_{n,m}(t) = \dfrac{\Omega_n^3}{2\omega_{pl}^2} u_{n,m} \chi_{zzz} E_P(t) \end{cases} \quad (13)$$

In the practically relevant case of a time-harmonic pump of the form $E_P(t) = A_P \cos(\Omega_P t)$, defining $d\Omega_{n,m} = \Omega_n^3 u_{n,m} \chi_{zzz} A_P / (2\omega_{pl}^2)$, equation (13) reduces to the well-known and extensively studied Mathieu equation [21, 22] with general solutions expressed in terms of Mathieu's Sine and Cosine functions as follow:

$$P_{n,m}(t) = a_{n,m} C\left(\frac{4\Omega_n^2 - \gamma^2}{\Omega_P^2}, -\frac{4\Omega_n d\Omega_n}{\Omega_P^2}, \frac{\Omega_P t}{2}\right) e^{-\frac{\gamma t}{2}} + $$
$$+ b_{n,m} S\left(\frac{4\Omega_n^2 - \gamma^2}{\Omega_P^2}, -\frac{4\Omega_n d\Omega_n}{\Omega_P^2}, \frac{\Omega_P t}{2}\right) e^{-\frac{\gamma t}{2}} \quad (14)$$

where the constants $a_{n,m}$ and $b_{n,m}$ are determined from the initial conditions. By expressing the Mathieu functions $F(a,q,x)$ in equation (14) in the Floquet form as the product of a periodic function $P(a,q,x)$ and a complex exponential, i.e. $F(a,q,x) = P(a,q,x) \exp(i\mu x)$, where $\mu$ is the characteristic Floquet exponent [23], the PPR threshold condition is obtained:

$$\text{Im}\left[\mu\left(\frac{4\Omega_n^2 - \gamma^2}{\Omega_P^2}, -\frac{4\Omega_n d\Omega_n}{\Omega_P^2}\right)\right] = \frac{\gamma}{\Omega_P} \quad (15)$$

The PPR threshold conditions for different collision frequencies $\gamma$ are shown in Figure 2 as a function of the normalized frequency modulation depth $d\Omega_n / \Omega_n$, and of the pump frequency $\Omega_P$. The lowest modulation threshold is obtained for $\Omega_P = 2\Omega_n$, when the pump field's frequency is twice the plasmonic mode frequency. For small losses, i.e. $\gamma \ll \omega_n$, and $\Omega_P = 2\Omega_n$ the threshold condition (15) can be expressed in terms of the $Q_n$ factor:

$$\frac{d\Omega_n}{\Omega_n} = \frac{2\gamma}{\Omega_n} = \frac{2}{Q_n} \quad (16)$$

From Figure 2 it is apparent that the PPR conditions can also be met, albeit with a higher threshold, when the pump frequency is detuned from the optimum value, which can be intuitively understood in terms of intervals of efficient energy transfer to the plasmonic mode spaced apart by longer intervals of attenuation.

As an example we apply the previous analysis to the specific case of a silver particle embedded in a 2-methyl-4-nitroanline (MNA) host [20]. MNA belongs to the m point group, with maximum nonlinear coefficient $\chi_{zzz} \sim 500 \, pm/V$. Considering the refractive index dispersion of MNA [24] and of silver [25], plasmonic resonances of high order occur around a wavelength of 450nm. For such system the calculated threshold pump intensity is $I_p \sim 1.4 [GW/cm^2]$. This intensity value, which is below the damage threshold of MNA [26], could be reached for instance with focused pulsed excimer lasers such as KrCl laser operating at the nearly optimum pump wavelength of 222nm. With a pulse duration of 100ps it is possible to achieve $\sim 10^4$ or more PPR amplification cycles. For particles of size comparable with the electron mean free path higher pump powers may be required, as a consequence of increased dissipation due to electronic scattering at the particle's surface [27]. Qualitatively similar results are expected for other variations of the configuration described so far, such as for instance a plasmonic shell surrounding a nonlinear core, or even larger particles at the onset of the Mie regime.

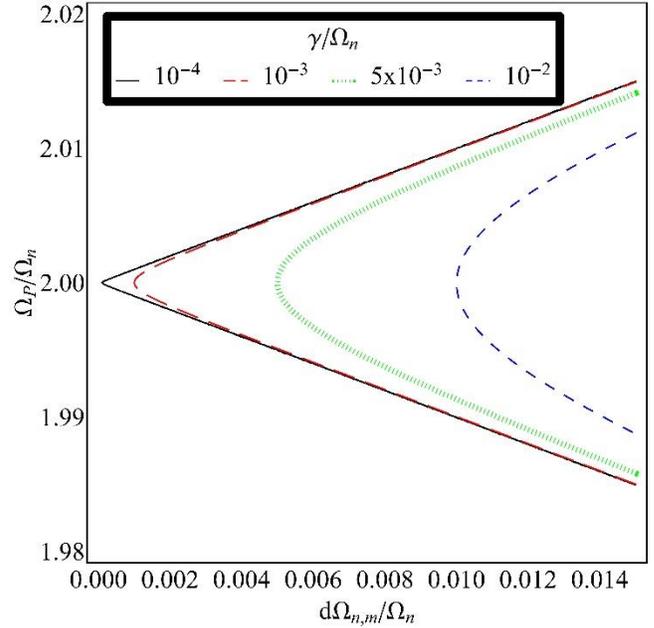

*Figure 2. PPR threshold conditions for various damping rates as a function of pump frequency and modulation amplitude.*

We would like to summarize here the salient features of PPR in comparison with conventional LSP resonance, as well as in comparison with other schemes of plasmonic amplification such as SPASER [17]. A fundamental characteristic of plasmonic parametric gain that emerges from the analysis above and in particular from equations (4), (7) and (9) is that a plasmonic mode of any order $(n,m)$ can undergo PPR and be amplified by a *spatially uniform* modulation of the background permittivity, provided that the temporal modulation profile is correct, and the appropriate threshold is exceeded. This is in stark contrast with conventional LSP resonance, which for a mode of order $(n,m)$ requires a driving field with a matching spatial profile – a condition nearly impossible to achieve in practice for high-order plasmonic modes of nanoparticles. For these reasons PPR is uniquely suitable to access plasmonic resonances of arbitrarily high order in deeply subwavelength structures.

A further distinction of PPR compared to LSPR must be noted: in order for a mode of order $(n,m)$ to undergo PPR it is necessary for such mode to be already oscillating in the system - however small its initial amplitude might be. In the linear regime, treating each mode as an independent and distinguishable harmonic oscillator in thermodynamic equilibrium with the background, such initial conditions can be easily determined according to a Planck distribution [28], implying that, at the very least, zero-point oscillations must exist, and therefore can be amplified by PPR. It is also worth pointing out that the correct initial phase relation between the pump and the plasmonic mode of interest is not a critical parameter, provided that the pump exceeds the PPR threshold. That is because PPR displays the important property of "phase-locking" that is common to all parametrically resonant systems [29], and that leads to a synchronization between the pump and the mode experiencing parametric gain [29].

Finally, when compared with other schemes of plasmonic amplification, such as SPASER[17] in which there is no coherence between plasmon and pump field, the coherent nature of the energy exchange leading PPR offers interesting advantages from the point of view of the detection of the occurrence of PPR. While the amplified modes are essentially non-radiative and difficult to detect directly, the instantaneous energy content of the PPR mode affects the pump field absorption and therefore the PPR amplification dynamics are expected to leave an imprint on the pump pulse temporal profile whereby its trailing edge would experience a larger attenuation – an easily measurable characteristic. For the same reason we may envision using collections of particles undergoing PPR for optical limiting applications.

In conclusion, we have introduced the concept and presented the theory of Plasmonic Parametric Resonance. Unlike conventional LSPR, all the plasmonic modes of a nanostructure, including the strongly subradiant ones, can be resonantly excited by spatially uniform optical pumping, provided that the corresponding threshold is exceeded. Accessing such high density of strongly localized states holds promise for enhancing nonlinear light-matter interaction at the nanoscale, and for the development of nonlinear optical metamaterials. Moreover, the coherent nature of the PPR process lends itself to simpler detection experiments compared with other plasmon amplification schemes.

**Acknowledgements**
We acknowledge AFOSR support through the 2016 Young Investigator Program award FA9550-16-1-0152.